# Structural Chirality of β-Mn

S. W. Lovesey[1,2]

[1]ISIS Facility, STFC, Didcot, Oxfordshire OX11 0QX, UK

[2]Diamond Light Source Ltd, Didcot, Oxfordshire OX11 0DE, UK

**Abstract** A theoretical study of Bragg diffraction by an enantiomorphic pair of structures should apply to patterns gathered on β-Mn using resonance enhanced x-ray diffraction. The chiral polymorph of manganese, and structurally related compounds, will be more convenient than trigonal crystals of low-quartz, tellurium and berlinite used in previous structural studies of this type, because cubic symmetry removes some complications of correction for absorption and macroscopic birefringence. Intensity of a Bragg spot engaged by circular polarization in the primary beam of photons $\Upsilon(h, k, l)$ is proposed as a chiral signature of the illuminated material. The partial intensity requires a knowledge of scattering amplitudes in all four channels of polarization, which are reported as functions of an azimuthal angle (rotation of the crystal about the axis of the reflection vector). Unlike trigonal chiral crystals, $\Upsilon(0, 0, l) = 0$ for β-Mn and it is $(h, k, 0)$, and symmetry related Bragg spots, that epitomize structural chirality. Specifically, reflection vectors $(h, k, 0)$ with odd Miller indices and the chiral axis of β-Mn mesh in terms of helicity, with the corresponding $\Upsilon(h, k, 0)$ equal in magnitude and opposite in sign for partners in the enantiomorphic pair. Dependence of $\Upsilon(h, k, 0)$ on the azimuthal angle does not mirror the dyad or tetrad axes of rotation symmetry in the cubic crystal structure.

## I. INTRODUCTION

The general public was made aware of the significance of handed forms of chemical compounds in the 1960s through devastating effects on human life caused by the medically approved consumption of racemic thalidomide by pregnant women. One handed form of thalidomide is a useful sedative while its opposite form is teratogenic [1]. A simple way to distinguish left- from right-handed forms is by passing visible light through them and measuring the plane of polarization of the transmitted light. The polarization plane of light gets rotated in a clockwise or counter-clockwise direction, and is used to distinguish the two forms. Optical activity is found in natural organic substances such as sugars, camphor, or tartaric acid, as well as in inorganic materials such as low-quartz and tellurium. Application of the technique in the wider domain is illustrated in a detective novel where murder is established after forensic scientists distinguish a synthetic poison (racemic) from a natural one (muscarine, handed) in a species of mushroom [2, 3]. Optical activity is a very small effect. One proposal to improve its efficacy in labelling handed forms of compounds exploits super-chiral beams of photons [4, 5]. Many optically active molecules or crystals have enantiomers, or stereoisomers, whose atomic configurations are exact mirror images of each other and are thus handed. This geometrical property of crystals and molecules is called chirality. Chiral substances possess a unique architecture such that, despite sharing identical molecular formulas, ion-to-ion linkages, and bonding distances, they cannot be superimposed. One chiral form does not match its mirror image (enantiomer). The occurrence of homochirality in amino acids and sugars is an essential enigma in biology, possibly related to the formation of life. There is no consensus on the mechanism that gives the homochirality of life at the present time [6].

Analysis of x-ray diffraction patterns reveals the absolute chirality of crystals in favourable cases [7]. However, single-crystal x-ray diffraction patterns remain essentially indistinguishable for enantiomorphs of chiral structures which contain only scattering ions of the same type, even when anomalous dispersion effects are taken into account. Higher-order harmonic generation and Kikuchi diffraction in electron scattering have recently been added to the list of techniques for the determination of chiral structures [8, 9]. Likewise, resonant x-ray Bragg diffraction is a proven technique by which to label chiral structures, and it is the central topic of the present study [10-15]. The technique is more selective than optical activity, which is allowed in 15 crystal classes, four of which are not enantiomeric. Alas, x-ray-absorption measurements are not forthcoming on structural chirality, for measurements involve forward scattering and present some averaged electronic information. Specifically, we illustrate the ability of Templeton and Templeton (T & T) scattering [16, 17] to illuminate the structural chirality of crystalline β-Mn through calculations of all scattering amplitudes [18, 19].

Manganese is known to be the most complex metallic element. The cubic β-Mn polymorph of interest does not order magnetically. Recent studies of the electronic properties of chiral structures [20-24] include β-Mn-type materials, e.g., Co-Zn-Mn alloys [25-28]. Yet, little attention has been given to structural chirality of the cubic form that we address [9]. The range of challenging properties found in chiral structures is wide, e.g., skyrmions [20, 26], current induced magnetization [21], gyrotropic effects [22], and topological conduction [24].

The β-Mn form contains 20 ions per unit cell in cubic enantiomorphs $P4_132$ and $P4_332$. The attendant crystallographic complexity sets it aside from low-quartz, tellurium and berlinite ($AlPO_4$) with enantiomorphs that belong to trigonal $P3_121$ (*right*-handed screw) and $P3_221$ (*left*-handed screw), and resonant ions occupy sites with multiplicity three [8, 10-14]. Diffraction patterns and azimuthal-angle scans (rotation of the diffracting crystal about the axis of the reflection vector) were measured with the energy of primary x-rays was tuned to K-edges (1s) of Si and Al, for example. Of 11 enantiomorphic pairs of space groups four have four-fold screw axes, namely, three tetragonal pairs $P4_1$ & $P4_3$, $P4_1 2 2$ & $P4_3 2 2$, $P4_1 2_1 2$ & $P4_3 2_1 2$, and the cited cubic structure for β-Mn. Inversion symmetry is absent in all four pairs while only $P4_1$ & $P4_3$ are polar structures. Scattering amplitudes for the polar enantiomorphs have one feature in common with the trigonal structures used by low-quartz, in that the chiral property is revealed when the reflection vector is aligned with the screw axis, and, also, measured azimuthal-angle scans on trigonal (0, 0, *l*) Bragg spots from low-quartz are three-fold periodic. We find this is not the case for β-Mn, and azimuthal angle scans on reflection vectors parallel to cell edges do not mesh with the four-fold screw axis.

Valence states that accept the photo-ejected electron, a few eV above the Fermi level, interact with neighbouring ions. In consequence, any corresponding electronic multipole is rotationally anisotropic with a symmetry corresponding to the site symmetry of the resonant ion. This anisotropy is most pronounced in the direct vicinity of an absorption edge whereas it is negligible far from the edges. Non-resonant ions can be neglected in calculations of forbidden reflection structure factors, to a good approximation. There are many reported examples of Bragg diffraction enhanced by absorption at the K-edge of a 3d transition ion. Results on haematite ($Fe^{3+}$, $3d^5$) reported by Finkelstein *et al*. [29] are thoroughly discussed by Cara and Thole [30], while diffraction patterns gathered at a later date reveal the material's chirality [31]. The time between the publications saw reports of diffraction patterns enhanced by nickel and vanadium K-edges [32-35]. A challenge posed by charge-orbital ordering in mixed valence perovskites was an early beneficiary of a strong resonance at the Mn K-edge [36, 37]. Likewise, for some manganese oxides, with Mn K edge measurements free of multiplet effects that complicate Mn L or K edge pre-peak spectra [38].

## II. MATERIAL PROPERTIES

β-Mn crystallizes in the enantiomorphic space-group pair $P4_132$ (No. 213) & $P4_332$ (No. 212) [18]. These are non-centrosymmetric structures that belong to the crystal class 432 (O), which is non-polar and not compatible with ferromagnetism. Mn ions occupy sites 12(d) and 8(c) with site symmetries $2_{yz}$ and $3_{xyz}$, respectively. Wyckoff groups are independent and the corresponding diffraction intensities add. Notably, two-fold and four-fold screw axes coincide. The crystal structures of β-Mn are depicted in Fig. 1. In the case of the β-Mn-type alloy $Co_{10}Zn_{10}$ cobalt ions uses 8(c) while both cobalt and zinc are in sites 12(d) [27]. The Bragg angle for β-Mn is estimated from $\sin(\theta) \approx 0.150 [h^2 + k^2 + l^2]^{1/2}$ based on a lattice constant $a \approx 6.29$ Å [18] and a value 6.537 keV for the energy of the Mn K-edge. The Bragg condition is not satisfied at $L_2$ ($\approx 0.649$ keV) or $L_3$ ($\approx 0.638$ keV) absorption edges.

Different sites in L-edge x-ray absorption spectroscopy (XAS) can be distinguished by energy only in favourable cases. For instance, the three Fe sites in magnetite can be separated, due to difference in valence state, point-group symmetry ($O_h$ and $T_d$), and in x-ray magnetic circular dichroism by opposite signs of $O_h$ and $T_d$ sites. It is an advantage that 3d electrons are localized so that the line width is narrow. In the case of β-Mn, likely the two sites are too similar in energy and line shape to separate in the L edge XAS. Also, the metallic line shape of Mn should be relatively quite broad.

Electronic properties are here expressed in terms of spherical multipoles $\langle O^K_Q \rangle$ of integer rank K with projections $-K \leq Q \leq K$ (Cartesian and spherical components of a dipole **R** = (x, y, z) are related by $x = (R_{-1} - R_{+1})/\sqrt{2}$, $y = i(R_{-1} + R_{+1})/\sqrt{2}$, $z = R_0$). The complex conjugate is defined as $\langle O^K_Q \rangle^* = (-1)^Q \langle O^K_{-Q} \rangle$, with a phase convention $\langle O^K_Q \rangle = [\langle O^K_Q \rangle' + i\langle O^K_Q \rangle'']$ for real and imaginary parts labelled by single and double primes, respectively. Multipoles are properties of the electronic ground state, and angular brackets $\langle ... \rangle$ denote the time-average, or expectation value, of the enclosed operator.

Implementation of sites symmetries is discussed in Appendix A. Quadrupoles $[\langle O^2_{+1} \rangle + \langle O^2_{-1} \rangle] = 2i\langle O^2_{+1} \rangle'' = 2i\langle O^2_{-1} \rangle''$ satisfy site symmetry $2_{yz}$, and they create diffraction at space-group forbidden reflections, e.g., (h, 0, 0) and (0, k, 0) with odd Miller indices. Likewise,

[⟨O³₊₂⟩ − ⟨O³₋₂⟩] = 2i⟨O³₊₂⟩″ = − 2i⟨O³₋₂⟩″ obeys site symmetry $3_{xyz}$, and T & T Bragg spots indexed by (h, 0, 0) are allowed for h = 2(2n +1), for example. This class of Bragg spots, due to ions at sites 8(c), do not differentiate between enantiomorphs. However, intensities of Bragg spots due to ions at sites 12(d) are different for space-groups No. 213 and No. 212.

### III. CHIRAL SIGNATURE

Scattered intensity picked out by circular polarization in the primary beam = $P_2 \Upsilon$ with [39],

$$\Upsilon = \{(\sigma'\pi)^*(\sigma'\sigma) + (\pi'\pi)^*(\pi'\sigma)\}'', \qquad (1)$$

and the Stokes parameter $P_2$ (a purely real pseudoscalar) measures helicity in the primary x-ray beam. Since intensity is a scalar quantity, $\Upsilon$ and $P_2$ must possess identical discrete symmetries, specifically, both scalars are time-even and parity-odd. In Eq. (1), and $(\pi'\sigma)$, for example, is the scattering amplitude for a rotated channel of polarization, cf. Fig. 2. Partial intensity $\Upsilon$ different from zero is a signature of a chiral motif of electronic multipoles, of course. Intensity of a Bragg spot in the rotated channel of polarization is proportional to $|(\pi'\sigma)|^2$, and likewise for unrotated channels of polarization.

By way of a relatively simple introductory example, we reproduce chiral signatures for the enantiomorphic space-group pair $P3_121$ (No. 152) and $P3_221$ (No. 154) appropriate for crystals of Te, $SiO_2$, and $AlPO_4$ mentioned in §I [10-14]. Sites 3(a) with symmetry $2_x$ are used by Te, Si and Al ions, and the crystal class is 32 ($D_3$). Axial (parity-even) multipoles are denoted by $\langle T^K_Q \rangle$. Two quadrupoles are engaged in Bragg intensities enhanced by an E1-E1 event, namely, $\langle T^2_{+2} \rangle'$ and $\langle T^2_{+1} \rangle''$. For the partial intensity of interest we find,

$$\Upsilon(0, 0, l) = \nu \, \langle T^2_{+2} \rangle' \, \{\sin(\theta) \, [1 + \sin^2(\theta)] \, \langle T^2_{+2} \rangle' - \cos^3(\theta) \cos(3\psi) \, \langle T^2_{+1} \rangle''\}. \qquad (2)$$

The azimuthal angle $\psi$ is rotation about a reflection vector (0, 0, l) parallel to the screw axis. A helicity index $\nu = +1$ for No. 152 with l = 1 and No. 154 with l = 2, while $\nu = −1$ for No. 152 with l = 2 and No. 154 with l = 1. These findings are consistent with experiments, to a very good approximation, and they illustrate the direct correlation between crystal chirality and the intensity of space-group forbidden Bragg spots picked out by circular polarization in the primary beam. In addition, dependence of $\Upsilon(0, 0, l)$ on the azimuthal angle faithfully mirrors the three-fold rotation symmetry. Because sites 3(a) are not centres of inversion symmetry polar (parity-odd) multipoles $\langle U^K_Q \rangle$ can be different from zero. Natural circular dichroism (NCD) is allowed and the signal is proportional to $\langle U^2_0 \rangle$ [40].

### IV. DIFFRACTION PATTERNS

An electronic structure factor for diffraction is $\Psi^K_Q = [\exp(i\boldsymbol{\kappa} \cdot \mathbf{d}) \, \langle O^K_Q \rangle_\mathbf{d}]$, where the Bragg wave vector $\boldsymbol{\kappa}$ is defined by integer Miller indices (h, k, l), and the implied sum in $\Psi^K_Q$ is over all 20 Mn sites in a unit cell. Bulk signals like NCD are proportional to $\Psi^K_Q$ evaluated for $\kappa = 0$. X-ray polarization vectors and the Bragg condition are depicted in Fig. 2. Extinction rules, or space-group-allowed reflections, possess $\Psi^0_0$ different from zero. Angular anisotropy in charge distributions make higher-order multipoles different from zero. In the present case, space-group forbidden charge-like T & T scattering is created in resonance enhanced Bragg diffraction by quadrupoles ($K = 2$) and, also, octupoles ($K = 3$).

Axial multipoles with K even can be observed using enhancement by E1-E1 and E2-E2 resonant events, for example, with the latter event relegated to Appendix B. On the other hand, even and odd rank polar multipoles are observed with enhancement by E1-E2 ($K = 1, 2, 3$) and E1-M1 ($K = 0, 1, 2$) resonant events. Convenient expressions for corresponding unit-cell structure factors are written in terms of even and odd combinations of $\Psi^K_Q$, defined by $\Psi^K_{\pm Q} = A^K_Q \pm B^K_Q$ [41]. Space groups of interest are generated by pure rotations. In consequence, we only need calculate $\Psi^K_Q$ for general $\langle O^K_Q \rangle$ and then specialize for a specific enhancement event.

*Sites* 8(c). Multipoles at sites 8(c) are allowed for rank K odd and K larger than 3. Thus, T & T scattering arises in diffraction enhanced by the E1-E2 parity-odd event. In the present case, diffraction is determined by $A^3_2$ and $B^3_2$ alone [41] and required scattering amplitudes are,

$$(\sigma'\sigma)_{12} = -\sin(\theta)\sin(2\psi) B^3_2, \quad (\pi'\pi)_{12} = \sin^2(\theta) (\sigma'\sigma)_{12}, \qquad (3)$$

$$(\pi'\sigma)_{12} = (i/2) \sin(2\theta) \cos(\psi) A^3_2 - \sin^2(\theta) \cos(2\psi) B^3_2.$$

The fourth amplitude $(\sigma'\pi)_{12}$ is obtained from $(\pi'\sigma)_{12}$ by a change in sign of $B^3_2$. Inserting results for the scattering amplitudes in Eq. (1) yields $\Upsilon_{12}(h, k, l) = 0$, because required products of amplitudes create a purely real quantity. Suffixes on scattering amplitudes and the chiral signature specify the nature of the resonant event. Later, the labelling scheme is extended to other events and their corresponding quantities. Recall that $|(\pi'\sigma)_{12}|^2$ is our prediction for intensity of a Bragg spot in the rotated channel of polarization, say. Absence of projections other than $Q = \pm 2$ in E1-E2 amplitudes is a direct consequence of operations needed to construct the unit cell. For sites 8(c) the operations are very simple rotations, accompanied by translations, that do no alter the magnitude of Q. They include, $4_z \langle O^K_Q \rangle = \exp(i\pi Q/2) \langle O^K_Q \rangle$, $2_y \langle O^K_Q \rangle = (-1)^{K+Q} \langle O^K_{-Q} \rangle$ and $2_{xy} \langle O^K_Q \rangle = (-1)^K \exp(i\pi Q/2) \langle O^K_{-Q} \rangle$.

Since $\langle U^3_{+2} \rangle'' = -\langle U^3_{-2} \rangle''$ it follows that $\Psi^3_{+2} = -\Psi^3_{-2}$. Let us consider the Bragg diffraction pattern indexed by $(h, k, 0)$. In this case, $A^3_2 = i \sin(2\delta) \Psi^3_{+2}$ and $B^3_2 = \cos(2\delta) \Psi^3_{+2}$, where $\delta$ is the angle of rotation in the plane of scattering that aligns $(h, k, 0)$ with the axis $-x$ in Fig. 2, e.g., $\cos(2\delta) = (h^2 - k^2)/(h^2 + k^2)$. The crystal c-axis is normal to the plane of scattering at $\psi = 0$. Scattering amplitudes vanishes for $h = k$ and conditions $h, k = 4n$ that define space-group allowed reflections. The mentioned findings follow immediately from the general result,

$$\Psi^3_{+2}(213) = 2i\langle U^3_{+2} \rangle'' \{[\exp(i\varphi h) + (-1)^k \exp(-i\varphi h)] [\exp(i\varphi k) + (-1)^{h+k} \exp(-i\varphi k)]$$

$$- \exp(i\pi(-h + k)/2) \, [\exp(i\varphi h) + (-1)^{h+k} \exp(-i\varphi h)] \, [\exp(i\varphi k) + (-1)^h \exp(-i\varphi k)]\}, \quad (4)$$

Here, $\varphi = 2\pi x_o$ where $x_o \approx 0.064$ is the general coordinate [18]. Eq. (4) is valid for space group No. 213. Multiplication of the minus sign in Eq. (4) by $(-1)^{h+k}$ yields the corresponding result for No. 212. We then find, $\Psi^3_{+2}(213) = \Psi^3_{+2}(212) = 0$ for $h = k = 0$. Evidently $\Psi^3_{+2}(h, k, 0) = -\exp(i\pi(-h + k)/2) \, \Psi^3_{+2}(k, h, 0)$, and $\Psi^3_{+2}(h, 0, 0)$ is different from zero for $h = 2(2n + 1)$.

*Sites* 12(d). To begin with, from Section II and Appendix A site symmetry $2_{yz}$ is consistent with quadrupoles (K = 2) and projections Q = ±1. Scattering amplitudes for E2-E2 are the subject of Appendix B, and we continue with a discussion of E1-E1 amplitudes.

The electronic structure factor $\Psi^2_Q$ admits projections Q = ±1 and ±2, since four operations among the 12(d) ions in the unit cell induce Q = ±2. One operation in question is $2_{xz}\langle O^2_{+2}\rangle = [\langle O^2_{+1}\rangle + \langle O^2_{-1}\rangle]/2$. The three remaining operations obey symbolic relations $2_{-xz} = 4_y = -4_y^{-1} = -2_{xz}$. Application of these findings to the E1-E1 event yields scattering amplitudes,

$$(\sigma'\sigma)_{11} = -i \sin(2\psi) \, A^2_1 - \sin^2(\psi) \, A^2_2,$$

$$(\pi'\pi)_{11} = -i \sin^2(\theta) \sin(2\psi) \, A^2_1 + [1 - \sin^2(\theta) \sin^2(\psi)] \, A^2_2,$$

$$(\pi'\sigma)_{11} = -\sin(\theta) \, [i \cos(2\psi) \, A^2_1 + (1/2) \sin(2\psi) \, A^2_2]$$

$$+ \cos(\theta) \, [-\cos(\psi) \, B^2_1 + i \sin(\psi) \, B^2_2], \quad (5)$$

and $(\sigma'\pi)_{11}$ is obtained from $(\pi'\sigma)_{11}$ through a change in sign of $A^2_Q$. The amplitudes admit even and odd harmonics of $\psi$. Indeed, only $(\sigma'\sigma)_{11}$, being proportional to $\sin(\psi)$, has a dominant $\psi$-dependence. From Eq. (5) we establish a chiral signature,

$$\Upsilon_{11} = -\sin(\theta) \, [1 - \cos^2(\theta) \sin^2(\psi)] \, (A^2_1 A^{2*}_2)' \quad (6)$$

$$+ \cos(\theta) \cos(\psi) \, [1 + \cos^2(\theta) \sin^2(\psi)] \, (A^2_2 B^{2*}_1)'' - \cos^3(\theta) \sin(\psi) \sin(2\psi) \, (A^2_1 B^{2*}_2)''.$$

Note the absence in $\Upsilon_{11}$ of contributions using products of $A^2_1$ and $B^2_1$ or $A^2_2$ and $B^2_2$. Axial quadrupoles $\langle T^2_Q\rangle$ in $A^2_Q$ and $B^2_Q$ apply to the resonance event of interest. Sum-rules for partner absorption edges exist, cf. Eq. (73) in Ref. [40]. Their content is trivial in the present case, however, since multipoles are evaluated for the Mn K-edge. A radial integral $\langle 1s|R|4p\rangle^2$ is a pre-factor in calculated intensities, to be replaced by $\langle 1s|R^2|3d\rangle^2$ in the E2-E2 event discussed in Appendix B. Foregoing expressions for scattering amplitudes and the chiral signature are valid for all Miller indices. Next, we use $l = 0$, and some values of $\Upsilon_{11}(h, k, 0)$ derived from Eq. (6) are displayed in Fig. 3.

Values of $A^2_Q$ and $B^2_Q$ are derived from $\Psi^2_{\pm 1} = \alpha \pm i\beta$ and $\Psi^2_{\pm 2} = \pm \gamma$ using the definition $A^2_Q + B^2_Q = \exp(iQ\delta)\,\Psi^2_Q$. Whereupon, $A^2_1 = (\alpha \cos\delta - \beta \sin\delta)$, $B^2_1 = i(\alpha \sin\delta + \beta \cos\delta)$, $A^2_2 = i\gamma \sin(2\delta)$, $B^2_2 = \gamma \cos(2\delta)$. For space group No. 213,

$$\alpha(213) = Z\,[\exp(i\pi h/4) - (-1)^k \exp(-i\pi h/4)]\,[\exp(i\chi k) + (-1)^{h+k} \exp(-i\chi k)],$$

(7)

$$\beta(213) = -2i\,Z\,\exp(i\pi h/2)\,\{[1 - (-1)^{h+k}]\sin(\chi h)\cos(\pi k/4) + [1 + (-1)^{h+k}]\cos(\chi h)\sin(\pi k/4)\},$$

$$\gamma(213) = -2i\,Z\,\sin(\chi h + \pi k/2)[\exp(i\chi k) - (-1)^h \exp(-i\chi k)].$$

where $Z = 2i\langle T^2_{+1}\rangle''$ and $\chi = 2\pi y_o$ with $y_o \approx 0.202$ [18]. Our results for $\Upsilon_{11}$ and $\Upsilon_{22}$ exploit the identity $(A^2_1 B^{2}_1{}^*)' = (\alpha\beta^*)'' = 0$ for all $h$, $k$.

We now use a generic notation $\Upsilon(h, k, l)$ for the chiral signature as results and discussions apply to both E1-E1 and E2-E2 events. There are no bulk signals, because all three quantities in Eq. (7) vanish for $h = k = 0$, a result in line with sites 8(c) with $\Upsilon(h, k, l) = 0$. In the case of sites 12(d), $\Upsilon(h, k, 0)$ is zero for $\gamma = 0$, because the chiral signature is provided by interference between multipoles with different projections. Other general properties of note are $\Upsilon(h, k, 0) = 0$ when $h, k = 4n$, meaning $\alpha = \beta = 0$, $\Upsilon(h, 0, 0) = \Upsilon(0, k, 0) = 0$, while $\Upsilon(h, k, 0) = -\Upsilon(-h, -k, 0)$. The latter identity follows from the corresponding phases acquired by $\alpha \to (-1)^{h+1}\alpha$, $\beta \to (-1)^{h+1}\beta$ and $\gamma \to (-1)^h\gamma$ under simultaneous changes in sign to $h, k$. With indices $h, k$ odd, $\alpha$ and $\beta$ are purely real and $\gamma$ is purely imaginary. The reverse over real and imaginary is true for $h = 4n$ and $k$ odd, and the corresponding $\Upsilon(h, k, 0)$ is also non-zero. Setting $h = k$ reveals $\alpha = \beta$ for space-group forbidden $h = (2m + 1)$ and $h = 2(2n + 1)$. In consequence, $A^2_1 = 0$ for Bragg spots $(h, h, 0)$, while $B^2_2 = 0$ because $\delta = 45°$.

After repeating for space group No. 212 the extensive algebra behind Eq. (7), one finds all-important relations,

$$\alpha(212) = \alpha(213),\quad \beta(212) = (-1)^{h+k}\beta(213),\text{ and } \gamma(212) = (-1)^h \gamma(213). \quad (8)$$

Consider the particular class of space-group forbidden reflections for which Miller indices are odd. In this case, the first two results in Eq. (8) tell us that linear combinations of $\alpha$ and $\beta$, which make up $A^2_1$ and $B^2_1$, are identical in the two enantiomorphs, while $A^2_2$ and $B^2_2$ derived from $\gamma$ alone simply have opposite signs in the two enantiomorphs. From these findings and Eqs. (6) and (B2) it follows that, for $h, k$ odd, chiral signatures $\Upsilon(h, k, 0)$ possess equal magnitudes and opposite signs for the two enantiomorphs. A feature of these reflections is that $\Upsilon(h, h, 0) \propto \cos(\psi)$, and $\Upsilon_{11}(3, 3, 0)$ and $\Upsilon_{22}(3, 3, 0)$ are included in Fig. 3. For the same conditions, the scattering amplitudes $(\pi'\sigma)_{11}$ and $(\pi'\sigma)_{22}$ are also proportional to $\cos(\psi)$ and vanishes for $\psi = 90°$ while $(\sigma'\sigma)$ and $(\pi'\pi)$ are different from zero. With $h = 2(2n + 1)$ the chiral signature $\Upsilon(h, h, 0) = 0$ for $\psi = 90°$, although diffraction patterns for the two enantiomorphs are identical when $h$ is even.

We close the discussion of diffraction enhanced by E1-E1 and E2-E2 events by considering Bragg spots (0, 0, *l*). Electronic structure factors $\Psi^2_{\pm1}$ ($\Psi^2_{\pm2}$) are different from zero for *l* odd (even), and *l* even are space-group forbidden. Specifically, for space group No. 213,

$$\Psi^2_{\pm1} = 4Z \sin(\chi l) [\pm 1 - (-1)^m], \ l = (2m + 1), \qquad (9)$$

$$\Psi^2_{\pm2} = \pm 4iZ (-1)^{n+1}, \ l = 2(2n + 1).$$

Scattering amplitudes are obtained using $B^2_1 = 4Z \sin(\chi l)$ and $B^2_2 = (-1)^m B^2_1$, and $A^2_1 = 4iZ (-1)^n$, with all other $A^2_Q$ and $B^2_Q$ zero. At the origin of the azimuthal angle-scan the a-axis is normal to the plane of scattering. Evidently, $\Upsilon(0, 0, l) = 0$ for *l* even ($A^2_2$ and all $B^2_Q$ zero) and *l* odd (all $A^2_Q$ zero).

To establish a difference between the two enantiomorphs revealed in (0, 0, *l*) Bragg spots, we examine the difference in structure factors $\Delta^K_Q = [\Psi^K_Q(213) - \Psi^K_Q(212)]$ that is particularly simple for 12(d), because contributions from sites generated by $4_y$ and $2_{xz} = -4_y$ cancel. One finds $\Delta^2_{+1} = \Delta^2_{-1} = -8Z \sin(\chi l) (-1)^m$ and $\Delta^2_{\pm2} = 0$ for $l = (2m+1)$. In consequence, E1-E1 and E2-E2 unit-cell scattering amplitudes only differ in one factor, namely, $B^2_2 = -\Delta^2_{+1}$.

## V. DISCUSSION

The cubic polymorph β-Mn is of particular interest because the chemical structure is chiral and it is used by materials with intriguing, and potentially useful, electronic properties [25-28, 42]. Even with the current interest, the structure type has not been investigated with an x-ray technique known to be well-suited for investigations of crystalline chirality. We refer to Bragg diffraction of x-rays with the benefit of intensity enhancement from an atomic resonance [10-15]. A comprehensive knowledge of β-Mn structural chirality would seem highly desirable in building a reliable model of the electronic properties.

Using the pair of enantiomorphic space groups $P4_132$ (No. 213) and $P4_332$ (No. 212) previously established for β-Mn from Bragg diffraction patterns generated by conventional Thomson scattering, we calculated all scattering amplitudes for resonant diffraction to be confronted with experimental results at a future date [18]. Some properties of the calculated amplitudes are not found in previous diffraction studies of materials with chemical structures described by the trigonal enantiomorphic pair $P3_121$ (No. 152) and $P3_221$ (No. 154), namely berlinite, tellurium and low quartz. The favourable Mn K-edge energy and β-Mn cell size give access to several Bragg spots. Many Bragg diffraction patterns exploiting 3d atomic resonances, including the Mn K-edge, have been reported in the past three decades.

While individual, polarization-dependent β-Mn scattering amplitudes merit testing as functions of azimuthal angle and Bragg angle, we encourage measurements of a partial intensity linked to circular polarization in the primary x-ray beam. Such intensity arises when helicity in the x-ray beam meshes with a chiral axis, and it is promoted as a signature of structural chirality. The partial intensity in question is denoted $\Upsilon(h, k, l)$, with $h, k, l$ integer Miller indices. For trigonal crystals and space-group forbidden Bragg spots, $\Upsilon$(No. 152) and $\Upsilon$(No. 154) are equal in magnitude are opposite in sign for a reflection vector parallel to the screw axis, i.e., the sign of $\Upsilon(0, 0, l)$ with $l \neq 3n$ is attached to an enantiomorph. Calculated for β-Mn, we find $\Upsilon(0, 0, l) = 0$, likewise the remaining symmetry related signatures. Whereas, $\Upsilon(h, k, 0)$ with odd Miller indices is shown to label β-Mn enantiomorphs.

The self-enantiomeric cubic space-group $P2_13$ (No. 198) is usually assigned to sodium bromate crystals. Templeton and Templeton confirmed the assignment, and measured weak Bragg reflections $(0, 0, l)$ forbidden by the systematic-absence rules for the $2_1$ screw-axis using signal enhancement from the bromine K-edge [16]. A $2_1$ screw-axis imposes the selection rule $(l + Q)$ even on bromine multipoles $\langle O^K_Q \rangle$. On top of this requirement, Br site symmetry $3_{xyz}$ in $P2_13$ demands projections Q even in an octupole, cf. Appendix B. The crystal class 23 (T) allows optical activity, as in the cases of the trigonal and cubic structures belonging to crystal classes 32 ($D_3$) and 432 (O), respectively. The natural circular dichroic signal is proportional to the polar quadrupole $\langle U^2_0 \rangle$ forbidden by site symmetry $3_{xyz}$. Likewise, symmetries of environments at sites 8(c) and 12(d) in $P3_121$ and $P3_221$ for β-Mn forbid $\langle U^2_0 \rangle$.

Returning to structural properties of sodium bromate, the self-enantiomorphic orthorhombic structure $P2_12_12_1$ (No. 19, crystal class $D_2$) accords with the measured diffraction pattern [16]. For Br ions in No. 19 occupy sites with no symmetry, so the screw-axis selection rule $(l + Q)$ even can accommodate reflections with $l$ odd studied by Templeton and Templeton. A model for sodium bromate based on $P2_12_12_1$ is completed by a Br axial quadrupole $\langle T^2_{+1} \rangle'$ with spatial symmetry $2_y$. Space-group No. 19 is chosen above two other candidates for a minimal model that similarly do not infringe translation symmetry in $P2_13$. A search of isotropy subgroups of $P2_13$ associated with zero propagation vectors identifies monoclinic $P2_1$ and trigonal $R3$. A monoclinic structure with Br ions in non-equivalent sites is not preferred above higher symmetry $P2_12_12_1$. Likewise for Br ions accommodated in a trigonal structure not possessed of the desired screw-axis selection rule.

## ACKNOWLEDGEMENTS

Dr K. S. Knight contributed to early stages of the reported study. Dr D. D. Khalyavin scrutinized aspects of the crystal physics and constructed Figs. 1 and 3. Professor G. van der Laan assisted with the atomic physics of Mn, and calculated dipole and quadrupole radial integrals reported in Appendix B. Dr Y. Tanaka provided valuable comments on the manuscript in its making.

## APPENDIX A: ROTATION MATRICES

The rotation $3_{xyz} = C_3[111]$ equates to a cyclic change of Cartesian coordinates represented by $(x, y, z) \rightarrow (z, x, y)$ and it is a symmetry of sites 8(c), together with $C_3[11{-}1] = (y, -z, -x)$, $C_3^2[11{-}1] = (-z, x, -y)$, etc., required by the cubic symmetry. For a multipole,

$$\langle O^K_Q \rangle_{zxy} = \exp(iq\beta)\, d^K_{Qq}(\beta)\, \langle O^K_q \rangle_{xyz}, \qquad (A1)$$

with an implied sum on projections labelled q, $d^K_{Qq}(\beta)$ is a standard Wigner rotation matrix [43], and the angle $\beta = \pi/2$. From (A1) it follows that $[\langle O^3_{+2} \rangle - \langle O^3_{-2} \rangle]_{zxy} = [\langle O^3_{+2} \rangle - \langle O^3_{-2} \rangle]_{xyz}$. Symmetry $2_{yz}$ of sites 12(d) leads to,

$$\langle O^K_Q \rangle_{-xzy} = \exp(i(Q+q)\beta)\, d^K_{Qq}(\beta)\, \langle O^K_q \rangle_{xyz}. \qquad (A2)$$

Identity (A2) is satisfied by $[\langle O^2_{+1} \rangle + \langle O^2_{-1} \rangle] = 2i\langle O^2_{+1} \rangle''$, where the equality follows from our definition $\langle O^K_Q \rangle^* = (-1)^Q \langle O^K_{-Q} \rangle$. Of two non-trivial rotations required to construct the 12(d) electronic structure factor, $2_{xz}(x, y, z) \rightarrow (z, -y, x)$ with,

$$2_{xz} \langle O^K_Q \rangle = (-1)^q\, d^K_{Qq}(\beta)\, \langle O^K_q \rangle, \qquad (A3)$$

while the second rotation $4_y(x, y, z) \rightarrow (z, y, -x)$ with,

$$4_y \langle O^K_Q \rangle = d^K_{Qq}(\beta)\, \langle O^K_q \rangle. \qquad (A4)$$

Evaluation of a sum on projections q in foregoing results is facilitated by use of an identity $d^K_{Q-q}(\beta) = (-1)^{K+Q}\, d^K_{Qq}(\beta)$ [43]. E.g., the result $4_y \langle O^2_Q \rangle = Z\, d^2_{Q1}(\beta)\,[1 + (-1)^Q]$ with $Z = 2i\langle O^2_{+1} \rangle''$ uses (A4) and site symmetry $2_{yz}$.

### APPENDIX B: E2-E2

Reduced radial matrix elements for Mn ($3d^5$) calculated from Cowan's program [44] are $\langle 1s|R|4p \rangle/a_o = -0.00354$ and $\langle 1s|R^2|3d \rangle/a_o^2 = 0.00095$, where $a_o$ is the Bohr radius. The dipole matrix element is small compared to its value for $L_{2,3}$ edges $\langle 2p|R|3d \rangle/a_o = -0.2062$, as expected. The four scattering amplitudes are,

$$(\sigma'\sigma)_{22} = (\pi'\pi)_{11},\ (\pi'\pi)_{22} = \cos(4\theta)\,(\sigma'\sigma)_{11}, \qquad (B1)$$

$$(\pi'\sigma)_{22} = \sin(3\theta)\,[i\cos(2\psi)\, A^2_1 + (1/2)\sin(2\psi)\, A^2_2]$$

$$+ \cos(3\theta)\,[\cos(\psi)\, B^2_1 - i\sin(\psi)\, B^2_2],$$

and $(\sigma'\pi)_{22}$ is obtained from $(\pi'\sigma)_{22}$ through a change in sign to $A^2_Q$. Notably, $(\pi'\sigma)_{22}$ can be derived from $(\pi'\sigma)_{11}$ by making replacements $\sin(\theta) \rightarrow -\sin(3\theta)$ and $\cos(\theta) \rightarrow -\cos(3\theta)$ in the latter. We go onto find a chiral signature,

$$\Upsilon_{22} = \sin(3\theta)\,(A^2_1 A^{2\,*}_2)'\{\cos(2\psi) + [\sin^2(\theta) + \cos(4\theta)]\sin^2(\psi)\}$$

(B2)

$$+ \cos(3\theta)\cos(\psi)\{[1 - \sin^2(\theta)\sin^2(\psi)](A^2_2 B^{2\,*}_1)'' + 2\sin^2(\theta)\sin^2(\psi)(A^2_1 B^{2\,*}_2)''\}$$

$$+ \cos(3\theta)\cos(4\theta)\cos(\psi)\sin^2(\psi)\{(A^2_2 B^{2\,*}_1)'' - 2(A^2_1 B^{2\,*}_2)''\}.$$

Some results for $\Upsilon_{22}(h, k, 0)$ are displayed in Fig. 3.

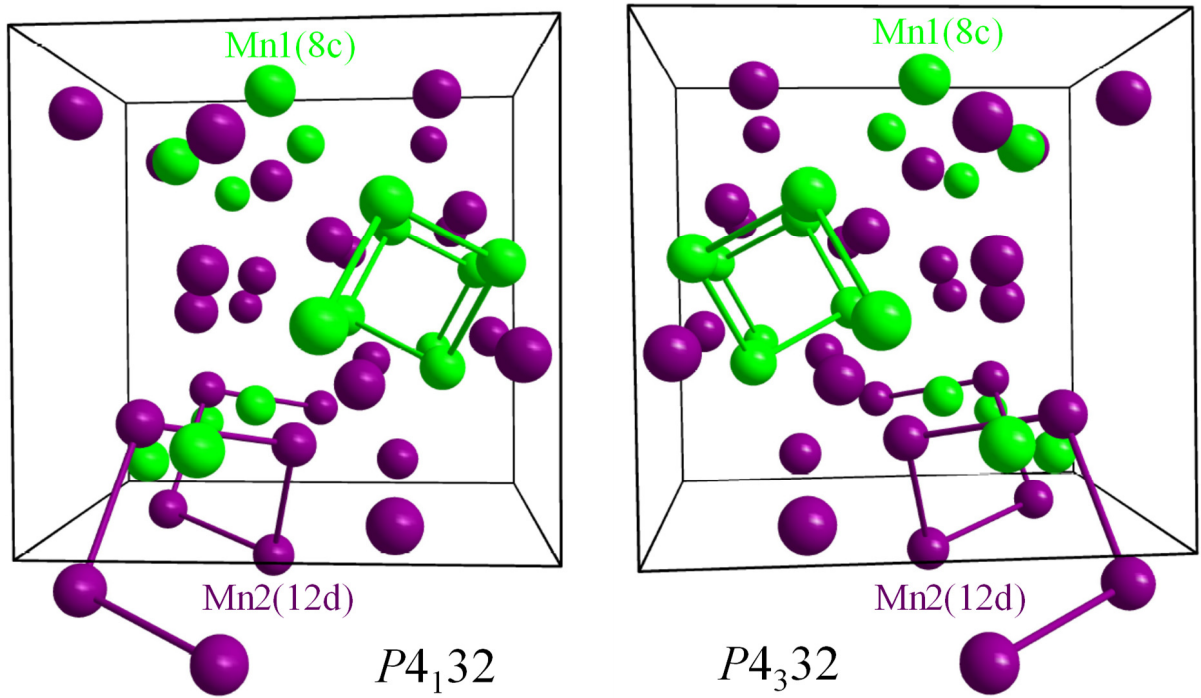

**FIG. 1.** Crystal structures of β-Mn enantiomorphs. See, also, Fig. 1 in Ref. [19].

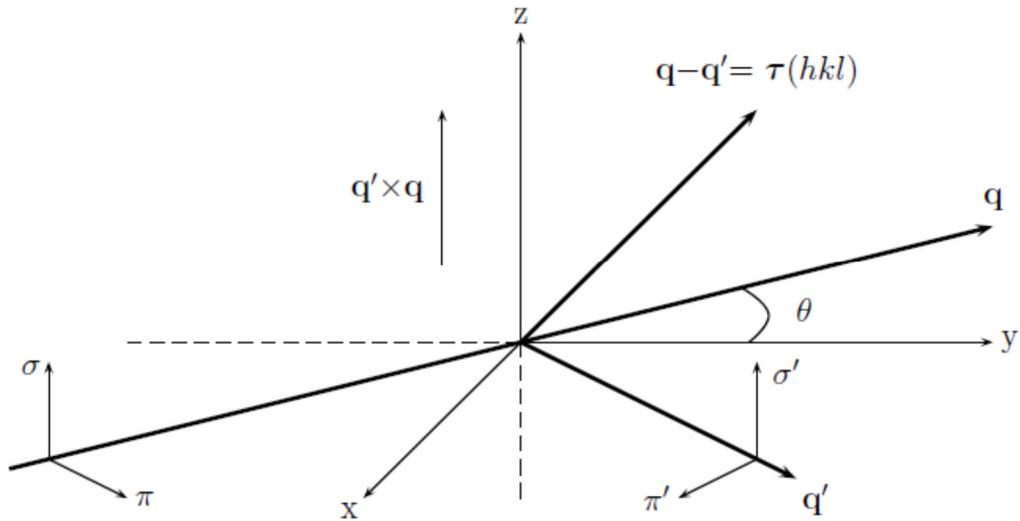

**FIG. 2.** X-ray diffraction coordinates. Primary ($\sigma$, $\pi$) and secondary ($\sigma'$, $\pi'$) states of polarization. Corresponding wave vectors **q** and **q'** subtend an angle $2\theta$, and the Bragg condition is met when $\boldsymbol{\kappa} = \mathbf{q} - \mathbf{q}'$ coincides with a reciprocal lattice vector $\boldsymbol{\tau}(hkl)$. Cell edges of a crystal and depicted Cartesian co-ordinates (x, y, z) coincide in the nominal setting of the crystal.

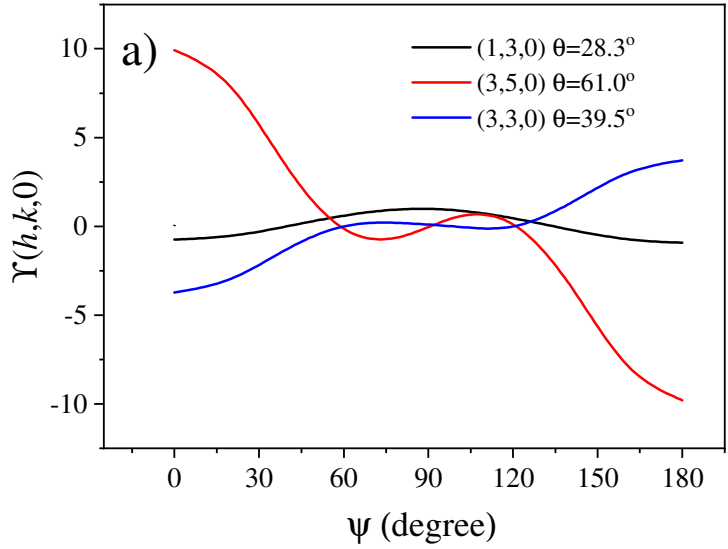

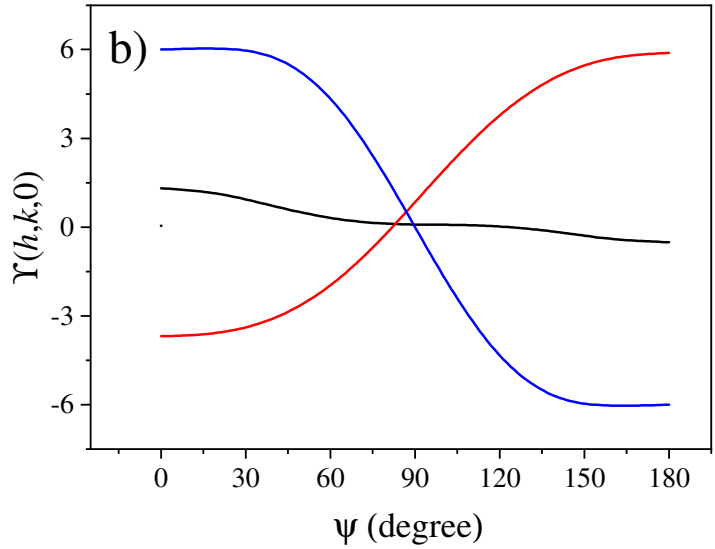

**FIG. 3**. The chiral signature $\Upsilon(h, k, 0) = \Upsilon(-h, -k, 0)$ of β-Mn for Miller indices $h$ and $k$ odd as a function of azimuthal angle $\psi$ in the range 0-180°. (a) E1-E1 event Eq. (6) and (b) E2-E2 event Eq. (B2) hold for P4$_1$32 (No. 213). Results for P4$_3$32 (No. 212) differ in sign alone from those displayed. Scans are symmetrical about $\psi$ =180°. Blue curve $\Upsilon(1, 3, 0)$, Bragg angle $\theta$ = 28.3°; orange curve $\Upsilon(3, 5, 0)$, $\theta$ = 61.0°; grey curve $\Upsilon(3, 3, 0)$, $\theta$ = 39.5°. $\Upsilon(h, k, 0)$ and $\Upsilon(k, h, 0)$ are related by a sign change and shift of 180°, with $\Upsilon(h, h, 0)$ zero at $\psi$ = 90°. Also, $\Upsilon(h, k, 0)$ and $\Upsilon(-h, k, 0)$ are related by a shift of 180° in the azimuthal angle.